\input harvmac.tex
\hfuzz 15pt
\input amssym.def
\input amssym.tex
\input epsf\def\tfig#1{{
\xdef#1{Fig.\thinspace\the\figno}}Fig.\thinspace\the\figno
\global\advance\figno by1}


\input epsf

%



\def\p{\partial}

\def\a{\alpha}
\def\b{\beta}
\def\g{\gamma}

\def\e{\epsilon}

\def\l{\lambda}

\def\G{\Gamma}
\def\D{\Delta}

\def\O{\Omega}

\def\ov{\over}

\def\no{\noindent}


 %

\def\[{\left[}
\def\]{\right]}
\def\({\left(}
\def\){\right)}
\def\<{\left\langle\,}
\def\>{\,\right\rangle}


\def\inv{^{-1}}

 \def\frac#1#2{ {{\textstyle{#1\over#2}}}}
\def\inv{^{\raise.15ex\hbox{${\scriptscriptstyle -}$}\kern-.05em 1}}

 \def\IP{\relax{\rm I\kern-.18em P}}


\def\hs{\hat{\s}}

\def\rb{ \noindent $\bullet$\ \ }

\def\dC{C\kern-6.5pt I}

              \def\CC{{\cal C}}
              
\def\CG{{\cal G}}              
              \def\CL{{\cal L}}

%



\chardef\tempcat=\the\catcode`\@ \catcode`\@=11
\def\cyracc{\def\u##1{\if \i##1\accent"24 i%
    \else \accent"24 ##1\fi }}
\newfam\cyrfam



\def\np#1#2#3{{Nucl. Phys.} {\bf B#1} (#2) #3}
\def\pl#1#2#3{{Phys. Lett. }{\bf B#1} (#2) #3}

\def\cmp#1#2#3{{Comm. Math. Phys.} {\bf #1} (#2) #3}
\def\mpl#1#2#3{{Mod. Phys. Lett. }{\bf #1} (#2) #3}
\def\ijmp#1#2#3{{Int. J. Mod. Phys.} {\bf #1} (#2) #3}
\def\lmp#1#2#3{{Lett. Math. Phys.} {\bf #1} (#2) #3}

\def\encadremath#1{\vbox{\hrule\hbox{\vrule\kern8pt\vbox{\kern8pt
 \hbox{$\displaystyle #1$}\kern8pt}
 \kern8pt\vrule}\hrule}}

\def\ee{\epsilon}
\def\tphi{\tilde\phi}
\def\hs{\hat{su}(2)}

\def\hepth#1{{arXiv:hep-th/}#1}

\def\np#1#2#3{{Nucl. Phys.} {\bf B#1} (#2) #3}
\def\pl#1#2#3{{Phys. Lett. }{\bf B#1} (#2) #3}

\def\cmp#1#2#3{{Comm. Math. Phys.} {\bf #1} (#2) #3}
\def\mpl#1#2#3{{Mod. Phys. Lett. }{\bf #1} (#2) #3}
\def\ijmp#1#2#3{{Int. J. Mod. Phys.} {\bf #1} (#2) #3}
\def\lmp#1#2#3{{Lett. Math. Phys.} {\bf #1} (#2) #3}

\def\jhep#1#2#3{{JHEP} {\bf #1} (#2) #3}




\lref\pogt{R. Poghossian, Two Dimensional Renormalization Group Flows in Next to Leading Order, JHEP {\bf 1401} (2014) 167; \hepth{1303.3015}.}
\lref\as{C. Ahn, M. Stanishkov, On the Renormalization Group Flow in Two Dimensional Superconformal Models, \np{885}{2014}713; \hepth{1404.7628}.}
\lref\gko{P. Goddard, A. Kent, D. Olive, Virasoro algebras and coset space models, \pl{152}{1985}88.}
\lref\myo{C. Crnkovic, G. Sotkov, M. Stanishkov, Renormalization group flow for general $SU(2)$ coset models, \pl{226}{1989}297.}
\lref\agt{L. Alday, D. Gaiotto, Y. Tachikawa, Liouville Correlation Functions from Four-dimensional Gauge Theories, \lmp{91}{2010}167; \hepth{0906.3219}.}
\lref\bfe{A. Belavin, B. Feigin, Super Liouville conformal blocks from $N=2$ $SU(2)$ quiver gauge theories, JHEP {\bf 1107} (2011) 079; \hepth{1105.5800}.}
\lref\bbb{A. Belavin, V. Belavin, M. Bershtein, Instantons and 2d Superconformal field theory, JHEP {\bf 1109} (2011) 117; \hepth{1106.4001}.}
\lref\abf{A. Belavin, M. Bershtein, B. Feigin, A. Litvinov, G. Tarnopolsky, Instanton moduli spaces and bases in coset conformal field theory, \cmp{319}{2013}269; \hepth{1111.2803}.}
\lref\afl{V. Alba, V. Fateev, A. Litvinov, G. Tarnopolsky, On combinatorial expansion of the conformal blocks arising from AGT conjecture, \lmp{98}{2011}33; \hepth{1012.1312}.}
\lref\at{M. Afimov, G. Tarnopolsky, Parafermionic Liouville field theory and instantons on ALE spaces, JHEP {\bf 1202} (2012) 036; \hepth{1110.5628}.}
\lref\myt{C. Crnkovic, R. Paunov, G. Sotkov, M. Stanishkov, Fusions of Conformal Models, \np{336}{1990}637.}
\lref\gai{D. Gaiotto, Domain Walls for Two-Dimensional Renormalization Group Flows, JHEP {\bf 1212} (2012) 103; \hepth{1201.0767}.}
\lref\ibr{I. Brunner, D. Roggenkamp, Defects and bulk perturbations of boundary Landau-Ginzburg orbifolds, \jhep{0804}{2008}001; \hepth{0712.0188}.}
\lref\ppt{A. Poghosyan, H. Poghosyan, RG domain wall for the $N=1$ minimal superconformal models, JHEP {\bf 1505} (2015) 043; \hepth{1412.6710}.}
\lref\mymy{M. Stanishkov, RG Domain Wall for the General $\hat{su}(2)$ Coset Models, \hepth{1606.03605}.}
\lref\ibc{I.Brunner, C. Schmidt-Colinet, Reflection and transmission of conformal perturbation defects, J.Phys. {\bf A49}(2016)195401; \hepth{1508.04350}.}
\lref\kmq{D. Kastor, E. Martinec, Z. Qiu, Current Algebra and Conformal Discrete Series, \pl{200}{1988}434.}
\lref\fr{F. Ravanini, An Infinite Class of New Conformal Field Theories With Extended Algebras, \mpl{3A}{1988}397.}
\lref\ag{P. Argyres, J. Grochocinski, S. Tye, Structure Constants of the Fractional Supersymmetry Chiral Algebras, \np{367}{1991}217; \hepth{9110052}.}
\lref\zam{A. Zamolodchikov, Renormalization Group and Perturbation Theory Near Fixed Points in Two Dimensional Field Theory, Sov. J. Nucl. Phys. {\bf 46} (1987) 1090.}
\lref\df{V. Dotsenko, V. Fateev, Operator Algebra of Two Dimensional Conformal Theories with Central  Charge $c<1$, \pl{154}{1985}291}
\lref\zp{A. Zamolodchikov, R. Poghossian, Operator algebra in two dimensional superconformal field theory, Sov. J. Nucl. Phys. {\bf 47}(1988)929.}
\lref\pogtri{R. Poghossian, Operator Algebra in Two Dimensional Conformal Quantum Field Theory Containing Spin $4/3$ Parafermionic Coserved Currents, \ijmp{A6}{1991}2005.}

\overfullrule=0pt
\Title{\vbox{\baselineskip12pt\hbox {}\hbox{}}} {\vbox{\centerline
 {Second Order RG Flow in}
  \vskip10pt
\centerline{General $\hs$ Coset Models }
  \vskip2pt
}} \centerline{ Marian Stanishkov\foot{marian@inrne.bas.bg } }

 \vskip 1cm

 \centerline{ \vbox{\baselineskip12pt\hbox {\it Institute for
Nuclear Research and Nuclear Energy,}
 }}
\centerline{ \vbox{\baselineskip12pt\hbox {\it Bulgarian Academy of Sciences, 1784 Sofia, Bulgaria}
 }}


\vskip 1.5cm

\centerline{ Abstract} \vskip.5cm \noindent \vbox{\baselineskip=11pt
We consider a RG flow in a general $\hat{su}(2)$ coset model perturbed by the least relevant field. The perturbing field as well as some particular fields of dimension close to one are constructed recursively in terms of lower level fields. Using this construction we obtain the structure constants and the four-point correlation functions in the leading order. This allows us to compute the mixing coefficients among the fields in the UV and the IR theory. It turns out that they are in agreement with those found recently using the domain wall construction up to this order.   }

\Date{}
\vfill \eject

%

\newsec {Introduction}

Recently there was some interest in the calculation of the matrix of anomalous dimensions and the corresponding mixing of certain fields in the two-dimensional CFT's perturbed by he least relevant field in the second order of the perturbation theory. This was done for the Virasoro theory in \pogt\ and then extended to the supersymmetric case in \as. Both theories are just  particular cases of more general $\hat{su}(2)$ coset models \gko. In this paper we extend the results of \pogt\ and \as\ to these models, denoted below as $M(k,l)$. The first order corrections were already obtained time ago in \myo. It was argued there that there exists an infrared (IR) fixed point of the renormalization group (RG) flow which coincides with the model $M(k-l,l)$. In the papers \pogt\ and \as\ the $\b$-function, the fixed point and the matrix of anomalous dimensions of certain fields were obtained up to the second order of the perturbation theory. Calculation up to the second order is always a challenge even in two dimensions. The problem is that one needs the corresponding four-point functions which are not known exactly. Fortunately, as explained in \pogt , one needs the value of these functions up to the zeroth order in the small parameter $\e={2\ov p+l}$.

Basic ingredients for the computation of the correlation function in two dimensions are the conformal blocks. These are quite complicated objects and a close form is not known. In the last years an exact relation between the latter and the instanton partition function of certain $N=2$ super YM theories in four dimensions was established by the so called AGT correspondence. This was done in \agt\ for the $l=1$ (Virasoro) and in \refs{\bfe ,\bbb} for the $l=2$ (supersymmetric) cases. More recently this correspondence was generalized to any $l$ in \refs{\abf ,\afl ,\at}.

In this paper we adopt another strategy. It was shown time ago in \myt\ that the structure constants and the conformal blocks for the general $\hat{su}(2)$ coset models $M(k,l)$ at some level $l$ can be obtained recursively from those of the lower levels or finally from the Virasoro minimal models themselves by certain projected tensor product. We use this construction here to define the perturbing field and the other fields in consideration. Following \myt\ we are able to compute the necessary structure constants and conformal blocks up to the desired order.

Another difficulty arises in the regularization of the integrals. In this paper we follow the regularization proposed in \pogt\ (see also \as).

There is an alternative approach to the calculation of the mixing matrix in the perturbed CFT models, the so called RG domain wall \gai\ (see also \ibr). $\quad$ As it was shown in \pogt\ for the Virasoro case and in \ppt\ for the supersymmetric extension, there is an agreement between the results obtained by such construction and the perturbative calculations up to the second order. Moreover, it was found in \refs{\pogt, \as} that the mixing matrix do not depend on $\e$ and is exactly the same in both theories. We showed recently in \mymy\ that this is the case also for the general $\hat{su}(2)$ coset models perturbed by the least relevant field in the first order of the perturbation theory. The goal of this paper is to check that result also in the second order.\foot{For a second order calculation of the so called reflection and transmission coefficients in general coset models we refer to \ibc.}

This paper is organized as follows.

In Section 2 we define the general $\hat{su}(2)$ coset models perturbed by the least relevant field $\tphi_{1,3}$ which is defined in terms of lower level fields. The basic ingredients necessary for the calculation in the second order of the perturbation theory are presented.

In Section 3 we give some more details needed for the computation of the conformal blocks. We explain, following \myt, how to construct the latter at level $l$ recursively in terms of lower level conformal blocks.

Section 4 is devoted to the calculation of the $\b$-function and the IR fixed point. It is confirmed that it coincides up to the second order with the model $M(k-l,l)$.

We define, similarly to the perturbing field itself, the fields with dimension close to 1 in section 5. Using the results for their correlation functions given in the Appendixes we compute the necessary two-point functions in the second order of the perturbation theory.

The matrix of anomalous dimensions for these fields is then computed in Section 6. It is in agreement with the first order calculation and the RG domain wall construction presented in \mymy. This is one of the main results of the present paper.

Finally, we present the calculation of the basic objects such as the structure constants and the correlation functions in the Appendixes. In Appendix A we explain how to use the construction of the fields and the conditions on their fusion rules in order to compute the corresponding structure constants. In Appendix B we present in some more details the calculation of the four-point function of the perturbing field. The calculation of these functions for the other fields is similar and is given in the last Appendix C.

\newsec{The theory}

Consider the $\hs$ coset model $M(k.l)$ \gko:
\eqn\coset{
{\hs_k\times \hs_l\over \hs_{k+l}}}
where $k$ and $l$ are integers and we assume that $k>l$ (we shall refer to $l$ as a level below). Here $\hs_k$ denotes the WZNW theory of level $k$. It is a conformal field theory (CFT) with the stress tensor $T_k$ given by the Sugawara construction. The corresponding central charge is $c_k={3k\ov k+2}$. The coset theory $M(k,l)$ \coset\ is then also a CFT, whose stress tensor is expressed through $T_k$  according to the construction: $T=T_k+T_l-T_{k+l}$ in obvious notations. The central charge of the corresponding Virasoro algebra is:
$$
c={3kl(k+l+4)\over (k+2)(l+2)(k+l+2)}={3l\over l+2}\(1-{2(l+2)\over (k+2)(k+l+2)}\).
$$
Here we will be interested in the "minimal models" with primary fields $\phi_{m,n}(l,p)$, $m$ and $n$ are integers. Their dimensions were computed in \kmq :
\eqn\dim{
\eqalign{
\D_{m,n}(l,p) &={((p+l)m-p n)^2-l^2\over 4lp(p+l)}+{s(l-s)\over 2l(l+2)},\cr
s &=|m-n|( mod (l)),\hskip1cm 0\le s\le l,\cr
&1\le m\le p-1, \hskip1cm 1\le n\le p+l-1}}
where we introduced ${\bf p=k+2}$ (note that we inverted $k$ and $l$  in the definition of the fields since we want to follow the notations of \refs{\myo ,\myt} ).

It is known \refs{\kmq ,\fr ,\ag } that the theory $M(k,l)$ possesses a symmetry generated by a "parafermionic current" $A(z)$ of dimension $\D_A={l+4\over l+2}$.
We don't need here the explicit construction of this current. We just mention that under this symmetry
the primary fields \dim\ are divided in sectors labeled by the integer $s$.

In this paper we prefer to use another description of the theory
$M(k,l)$ presented in \myt. It was shown there that this theory is not
independent but can be built out of products  of theories of lower
levels. Schematically this can be written as a recursion:
\eqn\proj{
M(1,l-1)\times M(k,l)={\bf P}(M(k,1)\times M(k+1,l-1))}
where ${\bf P}$ in the RHS is a specific projection. It allows the
multiplication of fields of the same internal indices and describes
primary and descendent fields (see \myt\ for more details).

In the following we will be interested in the CFT $M(k,l)$ perturbed by the least
relevant field. Such theory was described in \myo\ where the
$\b$-function and the fixed point were found. Recently, we computed also the mixing of certain fields along the corresponding RG flow in the first order of the perturbation theory  and compared that with the RG domain wall construction \mymy. The goal of this paper is to compute the mixing coefficients in the next to leading order.

Let us briefly sketch the constructions. The perturbed theory is
described by the Lagrangian:
$$
\CL(x)=\CL
_0(x)+\l \tilde\phi(x)
$$
where $\CL_0(x)$ describes the theory $M(k,l)$ itself. We
identify the field $\tilde\phi=\tilde\phi_{1,3}$ with the first descendent of
the corresponding primary field $\phi_{1,3}$  of dimension \dim with respect to the current
$A(z)$.  The dimension of this first descendent is \mymy:
\eqn\delt{
\D=\D_{1,3}+{l\over l+2}=1-{2\over p+l}=1-\e.}
In this paper we consider the case $p\rightarrow\infty$ and
assume that $\e={2\over p+l}\ll 1$ is a small parameter.

Following \myo\ we find it more convenient here to define the field $\tilde\phi_{1,3}$
alternatively in terms of lower level fields:
\eqn\field{
\tilde\phi_{1,3}(l,p)=a(l,p)\phi_{1,1}(1,p)\tilde\phi_{1,3}(l-1,p+1)+b(l,p)\phi_{1,3}(1,p)\phi_{3,3}(l-1,p+1).}
Here the field $\phi_{3,3}(l,p)$ is just a primary field constructed as:
\eqn\fitri{
\phi_{3,3}(l,p)=\phi_{3,3}(1,p)\phi_{3,3}(l-1,p+1)}
with dimension from \dim. It is straightforward to check that the field \field has a correct dimension \delt.
The coefficients $a(l,p)$ and $b(l,p)$ as well as the structure constants of the fields involved in the constructions \field\ and \fitri\ can be found by demanding the closure of the fusion rules \myo:
\eqn\frn{
\eqalign{
\tilde\phi_{1,3}(l,p)\tilde\phi_{1,3}(l,p) &=1+\CC_{(13)(13)}^{(13)}\tilde\phi_{1,3}(l,p)+\CC_{(13)(13)}^{(15)}(l,p)\tilde\phi_{1,5}(l,p),\cr
\phi_{3,3}(l,p)\phi_{3,3}(l,p) &=1+\CC_{(33)(33)}^{(13)}(l,p)\tilde\phi_{1,3}(l,p)+\CC_{(33)(33)}^{(33)}(l,p)\phi_{3,3}(l,p)+\cr
&+\CC_{(33)(33)}^{(15)}(l,p)\tilde\phi_{1,5}(l,p).}}
We found that
$$
a=\sqrt{{(l-1)(p-2)\ov l(p-1)}},\qquad b=\sqrt{{p-l-2\ov l(p-1)}},
$$
the structure constants are just a special case of those listed in the Appendix A.

In addition to \myo\ we introduced explicitly here the descendent field:
\eqn\five{
\tilde\phi_{1,5}(l,p)=x'(l,p)\phi_{1,1}(1,p)\tilde\phi_{1,5}(l-1,p+1)+y'(l,p)\phi_{1,3}(1,p)\tphi_{3,5}(l-1,p+1).}
of dimension $\tilde \D_{1,5}=2-{6\ov p+l}$.
In the same way as in \myo\ we find the coefficients and the structure constants involving this field:
\eqn\confive{
\eqalign{
x' &=\sqrt{{(l-2)(p-3)\ov l(p-1)}}, \qquad y'=\sqrt{{2(p+l-3)\ov l(p-1)}},\cr
\CC_{(33)(33)}^{(15)}(l,p) &=-\sqrt{{2l(l-1)\ov (p-2)(p-3)(p+l-3)(p+l-4)}} \tilde\CG_3(p+l-1),\cr
\CC_{(13)(13)}^{(15)}(l,p) &=(p+l-2)\sqrt{{2(l-1)(p-3)\ov l(p+l-3)(p+l-4)(p-2)}} \tilde\CG_3(p+l-1)}}
where the function $\tilde\CG_n(p+l-1)$ is defined in the Appendix A.

The mixing of the fields along the RG flow is connected to the two-point function. Up to the second order of the perturbation theory it is given by:
$$
\eqalign{
<\phi_1(x)\phi_2(0)>&=<\phi_1(x)\phi_2(0)>_0-\l\int <\phi_1(x)\phi_2(0)\tilde\phi(y)>_0 d^2y+\cr
&+{\l^2\ov 2}\int <\phi_1(x)\phi_2(0)\tilde\phi(x_1)\tilde\phi(x_2)>_0 d^2x_1 d^2x_2 +\ldots}
$$
where $\phi_1$, $\phi_2$ can be arbitrary fields of dimensions $\D_1$, $\D_2$.
The first order corrections were considered in \mymy, here we will focus on the second order. In doing that we follow closely \pogt (see also \as).

One can use the conformal transformation properties of the fields to bring the double integral to the form:
\eqn\bint{
\eqalign{
&\int <\phi_1(x)\phi_2(0)\tilde\phi(x_1)\tilde\phi(x_2)>_0 d^2x_1d^2x_2 =\cr
&=(x\bar x)^{2-\D_1-\D_2-2\D}\int I(x_1) <\tilde\phi(x_1)\phi_1(1)\phi_2(0)\tilde\phi(\infty)>_0 d^2x_1}}
where
$$
I(x)=\int |y|^{2(a-1)}|1-y|^{2(b-1)}|x-y|^{2c} d^2y
$$
and $a=2\e+\D_2-\D_1$,$b=2\e+\D_1-\D_2$, $c=-2\e$. It is well known that the integral for $I(x)$ can be expressed in terms of hypergeometric functions:
\eqn\ix{
\eqalign{
I(x)&={\pi\g(b)\g(a+c)\ov \g(a+b+c)}|F(1-a-b-c,-c,1-a-c,x)|^2+\cr
&+{\pi\g(1+c)\g(a)\ov \g(1+a+c)}|x^{a+c}F(a,1-b,1+a+c,x)|^2. }}

This form is useful for evaluating $I(x)$ near $x=0$.
Using the transformation properties of the hypergeometric functions, \ix\ can be rewritten as a function of $1-x$ and ${1\ov x}$ which is suitable for the investigation of $I(x)$ around the points $1$ and $\infty$, respectively.

It is clear that the integral \bint\ is singular. We follow the
regularization procedure proposed in \pogt . It was proposed there to cut discs in the two-dimensional surface of radius $r$
(${1\ov r}$) around singular points $0$, $1$ ($\infty$) with $0\ll r_0\ll r<1$, where $r_0$ is the ultraviolet cut-off.
The additional parameter $r$ is not physical and should not appear in the
final result. The region outside these discs, where the integration is well-defined, is called $\O_{r,r_0}$.
Near the
singular points one can use the OPE. The final result is a sum of all these contributions. It turns out however that
we count twice two lens-like regions around the point $1$ so we have
to subtract those integrals. We
refer to \pogt\ for the explicit formulas as well as a more
detailed explanation.

\newsec{Computation of the conformal blocks}

Let us consider the correlation function that enters the
integral \bint. The basic ingredients for the computation of the
four-point correlation functions are the conformal blocks. These are
quite complicated objects in general and closed formulae were not
known. Recently, it was argued that they coincide (up to factors)
with the instanton partition function of certain $N=2$ YM theories.
Here we adopt another strategy, similar to the calculation in $l=1$ \pogt\ and $l=2$ \as\ cases.
Namely, we find the expressions for
the conformal blocks up to a sufficiently high level in order to have a guess for
the limit $\e\rightarrow 0$.

According to the construction \proj\ any field $\phi_{m,n}(l,p)$ (or its descendent) can be expressed recursively as a product of lower level fields. Therefore the corresponding conformal blocks will be a product of lower level conformal blocks. Due to the RHS of \proj\ only certain products of conformal blocks will survive the projection ${\bf P}$. To be more explicit let us define the conformal block at level $l$ by
$$
F_l(r,s)=<\phi_{i_1,j_1}(x)\phi_{i_2,j_2}(0)|_{r,s}\phi_{i_3,j_3}(1)\phi_{i_4,j_4}(\infty)>_l
$$
where in the notation  we omitted the "external" fields and $r,s$ stands for the internal channel field $\phi_{r,s}$. The latter could be a primary field or a descendent. Which internal field can appear in the conformal block is defined by the fusion rules. The latter can be obtain recursively (see \myt\ for detailed explanation).

The conformal block is a chiral object, i.e. it depends only on the chiral coordinate $x$. It can be expanded as
\eqn\exp{
F(x)=x^{\D_{rs}-\D_{i_1j_1}-\D_{i_2j_2}}\sum_{N=0}^\infty x^N F_N}
where $N$ is called level (not to be confused with the level $l$ of $M(k,l)$) and we omitted the indexes.

In order to preserve the projection ${\bf P}$ in the intermediate channel, we allow only products of conformal blocks of the form:
\eqn\cbprod{
\eqalign{
&<\phi_{i_1,j_1}(x)\phi_{i_2,j_2}(0)|_{r,t}\phi_{i_3,j_3}(1)\phi_{i_4,j_4}(\infty)>_1\times\cr
&\times<\phi_{k_1,l_1}(x)\phi_{k_2,l_2}(0)|_{t,s}\phi_{k_3,l_3}(1)\phi_{k_4,l_4}(\infty)>_{1-1}\times\cr
&\times\sqrt{\CC_{(i_1j_1)(i_2j_2)}^{rt}\CC_{(i_3j_3)(i_4j_4)}^{rt}\CC_{(k_1l_1)(k_2l_2)}^{ts}\CC_{(k_3l_3)(k_4l_4)}^{ts}}.}}
Namely, only products of conformal blocks that involve the same internal indexes are allowed. Note that we included explicitly the corresponding structure constants. This is needed because they give different relative contribution on the subsequent levels in the expansion \exp. The overall constant will define the actual structure constant. Also, as explained in \myt, we take square roots of the structure constants because our considerations are chiral, i.e. depend only on the chiral coordinate $x$. Then, the true structure constant will be a square of the resulting one in \cbprod.

Actually, we consider below descendent fields which are some linear combinations like \field. Therefore we will have a linear combinations of products \cbprod. We give more details of the explicit construction of the conformal blocks in consideration in the Appendixes B and C.

The conformal blocks are in general quite complicated objects.
Fortunately, in view of the renormalization scheme and the
regularization of the integrals, we need to compute them here only
up to the zero-th order in $\e$. This simplifies significantly the
problem.

Once the conformal blocks are known, the correlation function of spinless fields for our $M(k,l)$ models is written as:
$$
\sum_{r,s} C_{rs}|F(r,s)|^2
$$
where the range of $(r,s)$ depends on the fusion rules and $C_{rs}$ is the
corresponding structure constant (we omitted the external indexes). The structure constants for the fields of interest are listed in Appendix A.

Our strategy here is to compute the conformal blocks recursively up to sufficiently high level. In addition we impose the condition of the crossing symmetry of the corresponding correlation function and the correct behaviour near the singular points 1 and $\infty$.

\newsec{$\beta$-function and fixed point}

For the computation of the $\b$-function up to the second order, we need
the four-point function of the perturbing field.
As explained in Appendix B  there are three ``channels" (or intermediate fields) in this
conformal block corresponding to the identity $\phi_{1,1}$,
$\tilde\phi_{1,5}$ and to $\tilde\phi$ itself. The explicit expression for
the correlation function is (B.8):
$$
\eqalign{
&<\tilde\phi(x)\tilde\phi(0)\tilde\phi(1)\tilde\phi(\infty)>=\cr
&= \left|{(1 - 2 x + ({5\ov 3}+{4\ov 3l}) x^2 - ({2\ov 3}+{4\ov 3l}) x^3 + {1\ov 3} x^4)\ov x^2 (1 - x)^2}\right|^2+{16\ov 3l^2}\left| {(1 - {3\ov 2} x + {(l+1)\ov 2} x^2 - {l\ov 4} x^3)\ov x (1 - x)^2}\right|^2+\cr
&+ {5\ov 9}\left({2(l-1)\ov l}\right)^2\left| {(1 - x +{l\ov 2(l-1)} x^2)\ov (1 - x)^2}\right|^2. }
$$
In the Appendix B we checked explicitly the crossing symmetry and the $x\rightarrow 1$ limit of this function.
In order to compute the $\beta$-function and the fixed point to the second order
we just have to integrate the above function.

The integration over the safe region far from the singularities
yields ($I(x)\sim {\pi\ov \epsilon}$):
$$
\eqalign{
&\int_{\Omega_{r,r_0}} I(x)<\tilde\phi(x)\tilde\phi(0)\tilde\phi(1)\tilde\phi(\infty)>d^2x\cr
&={(29 l^2-128 l) \pi^2\ov 24 \ee l^2} + {2 \pi^2\ov\ee r^2} + {\pi^2\ov 2 \ee r_0^2} -
{ 64 \pi^2 \log r\ov 3 \ee l^2} - {32 \pi^2 \log 2 r_0\ov 3 \ee l^2} }
$$
and we omitted the terms of order $r$ or $r_0/r$.

We have to subtract the integrals over the lens-like regions
since they  would be accounted twice.
Here is the result of that integration:
$$
{\pi^2 \ov \ee}
\(-{1\ov r^2}+{1\ov 2 r_0^2} +{1\ov 24}(29+{64\ov l})+{32\ov 3 l^2}\log{r\ov 2r_0}\).
$$

Next we have to compute the integrals near the singular points $0,1$ and $\infty$. For that purpose we can use the OPE of the
fields and take the appropriate limit of $I(x)$.
Near the point $0$ the relevant OPE is (by definition \frn):
$$
\tilde\phi(x)\tilde\phi(0)=(x\bar x)^{-2\Delta}(1+\ldots)
+ C_{(1,3)(1,3)}^{(1,3)}(x\bar x)^{-\Delta}(\tilde\phi(0)+\ldots).
$$
The channel $\tilde\phi_{1,5}$ gives after integration a term proportional
to $r/r_0$ which is negligible. The structure constant was computed in \myo.
Its value is
$$
 C_{(1,3)(1,3)}^{(1,3)}={4\ov l\sqrt 3} - 2 \sqrt 3 \ee
$$
to the first order in $\ee$. The value of $I(x)$ near $0$ can be found
by taking the limit in \ix\ written in terms of $1/x$ (explicit form
is given in \pogt).
Finally one gets:
$$
\int_{D_{r,0}\backslash D_{r_0,0}} I(x)<\tilde\phi(x)\tilde\phi(0)\tilde\phi(1)\tilde\phi(\infty)>d^2x
=-{\pi^2\ov r^2 \ee} + {32 \pi^2\ov
 3 l^2 \ee^2} -{32 \pi^2\ov l\ee} + {32\ov 3l^2} {\pi^2 \log r\ov\ee}
 $$
where the region of integration ${D_{r,0}\backslash D_{r_0,0}}$ is a ring with internal and external radiuses $r_0$ and $r$ respectively.
Since the integral near $1$ gives obviously the same result, we just need to add the above result twice.
To compute the integral near infinity, we use a relation
\eqn\infi{
<\phi_1(x)\phi_2(0)\phi_3(1)\phi_4(\infty)>=(x\bar x)^{-2\D_1}<\phi_1(1/x)\phi_4(0)\phi_3(1)\phi_2(\infty)> }
and $I(x)\sim {\pi\ov\ee}(x\bar x)^{-2\ee}$.
This gives
$$
\int_{D_{r,\infty}\backslash D_{r_0,\infty}} I(x)<\tphi(x)\tphi(0)\tphi(1)\tphi(\infty)>d^2x
=-{\pi^2\ov r^2 \ee} + {16 \pi^2\ov 3 l^2\ee^2}  - 16{ \pi^2\ov l\ee} + {32 \pi^2 \log r\ov 3l^2\ee}
 $$
where now ${D_{r,\infty}\backslash D_{r_0,\infty}}$ is a ring between ${1\ov r}$ and ${1\ov r_0}$.

Putting altogether, we obtain the finite part of the integral:
$$
{80\pi^2\ov 3l^2 \ee^2}-{88\pi^2\ov l\ee}.
$$
Notice that although the single integrals give different results, the final answer matches perfectly the $l=1$ \pogt\ and $l=2$ \as\ cases.
We want to mention also that we follow the renormalization scheme
proposed in \pogt. Therefore we already omitted the terms
proportional to $r_0^{4\ee-2}$ which could be canceled by an
appropriate counterterm in the action.

Taking into account also the first order term (proportional to
the above structure constant and computed in \mymy), we get the final
result (up to the second order) for the two-point function of the
perturbing field:
\eqn\twopt{
\eqalign{
G(x,\l)&=<\tphi (x)\tphi(0)>\cr
&=(x\bar x)^{-2+2\ee}\left[1-\l {4\pi\ov \sqrt 3}\({2\ov l\ee}-3\)(x\bar x)^\ee+{\l^2\ov 2}\({80\pi^2\ov 3l^2 \ee^2}-{88\pi^2\ov l\ee}\)(x\bar x)^{2\ee}
+\ldots\right]. }}

We now introduce a renormalized coupling constant $g$ and a renormalized field $\tphi^g=\p_g {\cal L}$ analogously to $\tphi=\p_{\l}{\cal L}$.
It is normalized by $<\tphi^g (1)\tphi^g(0)>=1$.
In this renormalization scheme the $\b$-function is given by \pogt:
$$
\beta(g)=\ee\l{\p g \ov\p\l}=\ee\l\sqrt{ G(1,\l)}
$$
where $G(1,\l)$ is given by \twopt\ with $x=1$.
One can invert this and compute the bare coupling constant and the $\beta$-function in terms of $g$:
\eqn\bare{\eqalign{
\l&=g+g^2{\pi\ov \sqrt 3}\left({2\ov l\ee}-3\right)+g^3{\pi^2\ov 3}\left({4\ov l^2\ee^2}-{10\ov l\ee}\right)+{\cal O}(g^4),\cr
\beta(g)&=\e g-g^2{\pi\ov\sqrt 3}({2\ov l}-3\e)-{4\pi^2\ov 3l}g^3+{\cal O}(g^4).}}
In this calculations, we keep only the relevant terms by assuming the
coupling constant $\l$ (and $g$) to be order of ${\cal O}(\e)$.

A non-trivial IR fixed point occurs at the zero of the $\beta$-function
\eqn\fx{
g^*={l\sqrt{3}\ov 2\pi}\e(1+{l\ov 2}\e).}
It corresponds to the IR CFT  $M(k-l,l)$ as can be seen from the central charge difference:
$$
c^*-c=-{4(l+2)\ov l}\pi^2\int_0^{g^*}\beta(g)d g=-l(1+{l\ov 2})\e^3-{3l^2\ov 4}(l+2)\e^4+{\cal O}(\e^5).
$$
The anomalous dimension of the perturbing field becomes
$$
\D^*=1-\p_g\beta(g)|_{g^*}=1+\e+l\e^2+{\cal O}(\e^3)
$$
which matches with that of the field $\phi_{3,1}(l,p-l)$ of $M(k-l,l)$ (defined in \mymy, see also the next Chapter).

\newsec{ Mixing of the fields }

The fields $\tilde\phi_{n,n\pm 2}$ defined recursively by (see \mymy):
\eqn\defn{
\eqalign{
\tilde\phi_{n,n+2}(l,p)&=x\phi_{n,n}(1,p)\tilde\phi_{n,n+2}(l-1,p+1)+y\phi_{n,n+2}(1,p)\phi_{n+2,n+2}(l-1,p+1),\cr
\tilde\phi_{n,n-2}(l,p)&=\tilde x\phi_{n,n}(1,p)\tilde\phi_{n,n-2}(l-1,p+1)+\tilde y\phi_{n,n-2}(1,p)\phi_{n-2,n-2}(l-1,p+1)}}
(where $x$, $\tilde x$ and $y$, $\tilde y$ are at $(l,p)$) and the derivative $\partial\phi_{n,n}$ of the primary field
\eqn\defnn{
\phi_{n,n}(l,p)=\phi_{n,n}(1,p)\phi_{n,n}(l-1,p+1).}
have dimensions close to $1$
\eqn\dimn{
\eqalign{
\tilde\D_{n,n\pm 2} &=1+{n^2-1\ov 4p}-{(2\pm n)^2-1\ov 4(p+l)}=1-{1\pm n\ov 2}\e+O(\e^2),\cr
1+\D_{n,n} &=1+{n^2-1\ov 4p}-{n^2-1\ov 4(p+l)}=1+{(n^2-1)l\ov 16}\e^2+O(\e^3)}}
and belong to the zero charge sector of the current $A(z)$. Together with the fusion rules which can be obtained recursively \myt\ this suggests that they mix along the RG-trajectory. We want to compute the matrix of anomalous dimensions and the corresponding mixing matrix of these fields. For that purpose we compute their two-point functions up to second order and the corresponding integrals.

\subsec{Function $<\tphi_{n,n+2}(1)\tphi_{n,n+2}(0)>$ }

The corresponding function in the second order of the perturbation theory can be found in Appendix C (C.1).
After transformation $x\rightarrow 1/x$ it becomes:
$$
\eqalign{
&<\tilde\phi(x)\tilde\phi_{n,n+2}(0)\tilde\phi_{n,n+2}(1)\tilde\phi(\infty)>=
\left|{(l - (2l+4) x + (5l+4) x^2 - 6l x^3 +3l x^4)\ov 3l x^2 (1 - x)^2}\right|^2+\cr
&+{8(n+3)\ov 3l^2(n+1)}\left| {(l -2(l+1) x +6 x^2 -4x^3)\ov 4 x^2 (1 - x)^2}\right|^2+\cr
&+ \left({2(l-1)\ov l}\right)^2{(n+3)(n+4)\ov 18 n(n+1)}\left| {(l +2(1-l) x +2(l-1) x^2)\ov 2(l-1)x^2 (1 - x)^2}\right|^2. }
$$

The integration of this function is very similar to that we did in the case of the computation of the $\b$-function. It goes along the same lines of the $l=1$  and $l=2$ cases so we do not present here the detailed calculation. The only difference is in the structure constants needed in the OPE's around $0$, $1$ and $\infty$. They were computed in \mymy\ and are given in the Appendix A:
\eqn\str{\eqalign{
(C_{(13)(nn+2)}^{(nn+2)})^2&={4(n+3)^2\ov 3l^2 (n+1)^2}-{4 (n+2) (n+3)^2 \e\ov 3 l(n+1)^2}+O(\e^2),\cr
(C_{(13)(nn+2)}^{(nn)})^2&={n+2\ov 3n}+O(\e^2) .\ } }

The final result of the integration is:
$$
{8\pi^2 (20 + 143 n + 121 n^2 + 33 n^3 + 3 n^4)\ov
 3l^2 n ( n+1) (n+3)^2 \e^2}-
{4\pi^2(n+5) (8 + 151 n + 143 n^2 + 45 n^3 + 5 n^4) \ov 3l n (n+1) (n+3)^2 \e}.
$$
This is in perfect agreement with $l=1$ and $l=2$ cases.

\subsec{Function $<\tphi_{n,n+2}(1)\tphi_{n,n-2}(0)>$ }

The relevant four-point function in this case in the zeroth order of $\e$ is given by (C.3). Transforming $x\rightarrow
{1\ov x}$ and  using \infi, one obtains:
$$
<\tphi(x)\tphi_{n,n+2}(1)\tphi_{n,n-2}(0)\tphi(\infty)>={1\ov 3} \sqrt{{(n^2-4)\ov n^2}}\left|{ 1\ov l x^2(1 - x)^2 }(l -2(l-1) x +2(l-1) x^2)\right|^2.
$$

Again, the integration over the safe region and lens-like region is very similar to $l=1$ and $l=2$ cases. The same is true also for the singular points where we have to take the structure constant:
$$
C_{(13)(nn-2)}^{(nn)}={n-2\ov 3n}+O(\e^2).
$$
Collecting all the integrals leads to the final result:
$$
{320 (1 - l \e) \pi^2\ov 3 l^2\e^2 n (n^2-9) \sqrt{n^2-4}}
$$
which again matches with Virasoro and superconformal cases.

\subsec{Function $<\phi_{n,n}(1)\tphi_{n,n+2}(0)>$ }

The four point function differs only in the structure constant (C.4):
$$
<\tphi(x)\phi_{n,n}(1)\tphi_{n,n+2}(0)\tphi(\infty)>={4\ov 3l}\sqrt{{n+2\ov n}}|x|^{-2}.
$$

Therefore the calculations are exactly the same. Also, the necessary structure constants for the calculation around singular points were already presented above. This leads to a final result:
$$
{4 (n-1)\pi^2\ov 3 l(n+3) (n+5)}\sqrt{{n+2\ov n}}\left[-22 -6 n + \e(-2 (n+5) (3 n+11)+l(46+n(n+15)))\right]
$$
which generalizes \pogt\ and \as.

\subsec{Function $<\phi_{n,n}(1)\phi_{n,n}(0)>$ }

Finally, we need the function $<\tphi(x)\phi_{n,n}(1)\phi_{n,n}(0)\tphi(\infty)>$.
As it is shown in Appendix C this function happens to coincide exactly with the one found in \pogt\ and \as and is given explicitly by (C.5).
Therefore almost all integrals are the same. The only exception  is
the integral around $\infty$ due to the different structure constants:
$$
C_{(13)(nn)}^{(nn)}C_{(13)(13)}^{(13)}={(n^2-1)  \e^2\ov 6  }(1-(l-2)\e).
$$
With this, the result is
$$
{(n^2-1)\pi^2 \ov 12}(2 +(8-3l) \e).
$$

Since the dimension of the field $\phi_{n,n}$ is close to zero, it doesn't mix with other fields.
Therefore, we need to compute only its anomalous dimension.
Taking into account also the first order
contribution \mymy, the final result for the two-point function is:
$$
\eqalign{ G_n(x,\l)=<\phi_{n,n}(x)\phi_{n,n}(0)>&=(x\bar
x)^{-2\D_{n,n}}\left[1 - \l \left({\sqrt 3 l\pi\ov 24} (n^2-1) \e(2+(l+4)\e) \right)(x\bar x)^\e\right.\cr
&+\left.{\l^2\ov 2} \left({\pi^2\ov 12} (2 +(8-3l)  \e) (n^2-1)\right)(x\bar x)^{2\e}+...\right].}
$$

Computation of the anomalous dimension goes in exactly the same way as for the perturbing field:
$$
\eqalign{
\D_{n,n}^g &=\D_{n,n}-{\e\l\ov 2}\p_\l G_n(1,\l)=\cr
&=\D_{n,n}+{\sqrt 3\pi g l\ov 48} \e^2 (2 + (l+4) \e) (n^2-1)+{\pi^2 g^2\ov 24} \e^2(l-4) (n^2-1)}
$$
where we again kept the appropriate terms of order $\e\sim g$. Then, at the fixed point \fx, this becomes
$$
\D_{n,n}^{g^*}=
{(n^2-1)l(4 \e^2 + 6l \e^3 + 7l^2 \e^4+...)\ov 64}
$$
which coincides up to the desired order with the dimension of the field $\phi_{n,n}(l,p-l)$
of the model $M(k-l,l)$.

\newsec{Matrix of anomalous dimensions}

Let us describe briefly the renormalization scheme of \pogt which
is a slight modification of the original one \zam\ due to Zamolodchikov.
We introduce renormalized fields $\phi^g_\a$ which are expressed through the bare ones by:
\eqn\defb{
\phi^g_\a=B_{\a\b}(\l)\phi_\b}
(here $\phi$ could be a primary or a descendent field).
The two-point functions of the renormalized fields
\eqn\norm{
G_{\a\b}^g(x)=<\phi_\a^g(x)\phi_\b^g(0)>,\quad G_{\a\b}^g(1)=\delta_{\a\b} }
satisfy the Callan-Symanzik equation:
$$
(x\p_x-\b(g)\p_g)G_{\a\b}^g+\sum_{\rho=1}^2(\G_{\a\rho}G_{\rho\b}^g+\G_{\b\rho}G_{\a\rho}^g)=0.
$$
The matrix of anomalous dimensions $\Gamma$ that appears above is given by
\eqn\ano{ \G=B\hat\D
B^{-1}-\e\l B\p_\l B^{-1} }
where $\hat\D=diag(\D_1,\D_2)$ is a
diagonal matrix of the bare dimensions.
The matrix $B$, as defined in \defb, is
computed from the matrix of the bare two-point functions we computed,
using the normalization condition \norm\ and requiring the matrix
$\G$ to be symmetric. Exact formulas can be found in \pogt.

We computed above some of the entries of the $3\times 3$ matrix
of two-point functions in the second order. This matrix is obviously
symmetric. It turns out also that the remaining functions
$<\tphi_{n,n-2}(1)\tphi_{n,n-2}(0)>$ and
$<\phi_{n,n}(1)\tphi_{n,n-2}(0)>$ can be obtained from the computed
ones $<\tphi_{n,n+2}(1)\tphi_{n,n+2}(0)>$ and
$<\phi_{n,n}(1)\tphi_{n,n+2}(0)>$ by just taking $n\rightarrow -n$.

Let us combine the fields in consideration in a vector with components:
$$
\eqalign{ \phi_1=\tphi_{n,n+2},\quad
\phi_2=(2\D_{n,n}(2\D_{n,n}+1))^{-1}\p\bar\p \phi_{n,n},\quad
\phi_3=\tphi_{n,n-2}. }
$$
The field $\phi_2$ is normalized so that its bare two-point function is $1$. It is straightforward to
modify the functions involving $\phi_2$ taking into account the derivatives and the normalization.

We can write the matrix of the two-point functions up to the second
order in the perturbation expansion as:
$$
\eqalign{
G_{\a,\b}(x,\l)&=<\phi_\a(x)\phi_\b(0)>=
(x\bar x)^{-\D_\a-\D_\b}\left[\delta_{\a,\b}-\l C^{(1)}_{\a,\b}(x\bar x)^{\e}+{\l^2\ov 2}C^{(2)}_{\a,\b}(x\bar x)^{2\e}+...\right].}
$$

The two-point functions in the first order are proportional to the
structure constants: \eqn\cfir{
 C^{(1)}_{\a,\b}=C_{(1,3)(\a)(\b)}{\pi \g(\e+\D_\a-\D_\b)\g(\e-\D_\a+\D_\b)\ov
 \g(2\e)}}
which are symmetric.

Collecting all the dimensions and structure
constants, we get
$$
\eqalign{
C^{(1)}_{1,1}&=-{2 (n+3) (-2 +l\e(n+2) ) \pi\ov \sqrt 3 l\e (n+1)},\quad
C^{(1)}_{1,2}={8 (-2 +l \e) \sqrt{{n+2\ov n}} \pi\ov \sqrt 3 l \e (n+1) (n+3)},\quad
C^{(1)}_{1,3}=0,\cr
C^{(1)}_{2,2}&={16  \pi\ov \sqrt 3  l(n^2-1) \e} - {4  (n^2+1) \pi\ov
 \sqrt 3 (n^2-1) },\quad
C^{(1)}_{2,3}={8 (-2 + l\e) \sqrt{{n-2\ov n}} \pi\ov \sqrt 3 l\e (n-3) (n-1)},\cr
C^{(1)}_{3,3}&={-2 (n-3) (-2 +  l\e(2-n)) \pi\ov \sqrt 3 l\e (n-1)}}
$$
for the first order, and
$$
\eqalign{
C^{(2)}_{1,1}&={8 (20 + 143 n + 121 n^2 + 33 n^3 + 3 n^4) \pi^2\ov
 3l^2 n (n+1) (n+3)^2 \e^2}-\cr
 &-{4 (n+5) (8 + 151 n + 143 n^2 + 45 n^3 + 5 n^4) \pi^2\ov
 3l n (n+1) (n+3)^2 \e},\cr
C^{(2)}_{1,2}&=-{64 \sqrt{{n+2\ov n}} (3 n+11) \pi^2\ov
  3l^2 (n+1) (n+3) (n+5) \e^2} +{ 32 \sqrt{{n+2\ov n}} (57 + 18 n + n^2) \pi^2\ov
 3l (n+1) (n+3)(n+5) \e },\cr
C^{(2)}_{1,3}&={320 (1 - l \e) \pi^2\ov 3l^2 \e^2 n (n^2-9) \sqrt{n^2-4}},\cr
C^{(2)}_{2,2}&={128 \pi^2\ov 3l^2 (n^2-1) \e^2} - {16 (n^2+19) \pi^2\ov
 3l (n^2-1) \e},\cr
C^{(2)}_{2,3}&=-{64 \sqrt{{n-2\ov n}} (3 n-11) \pi^2\ov
  3l^2 n-1) (n-3) (n-5)  \e^2} -  {32 \sqrt{{n-2\ov n}} (57 - 18 n + n^2) \pi^2\ov
 3l (n-1) (n-3) (n-5)  \e},\cr
C^{(2)}_{3,3}&= -{8 (-20 + 143 n - 121 n^2 + 33 n^3 - 3 n^4) \pi^2\ov
  3l^2 n (n-1) (n-3)^2 \e^2} +\cr
  &+ {4 (n-5) (8 - 151 n + 143 n^2 - 45 n^3 + 5 n^4) \pi^2\ov
 3l n (n-1) (n-3)^2 \e} }
$$
for the second one.

Now we can apply the renormalization procedure of \pogt\ and obtain
the matrix of anomalous dimensions \ano. The bare coupling constant $\l$ is
expressed through $g$ by \bare\ and the bare dimensions, up to order
$\e^2$. The results are:
$$
\eqalign{
\G_{1,1}&=\D_1-{(n+3) (-2 +l \e (2 + n)) \pi g\ov \sqrt 3 l(n+1)}+{8 g^2 \pi^2(n+2)\ov 3l (n+1)},\cr
\G_{1,2}&=\G_{2,1}=-{(-2 + l\e) (n-1) \sqrt{{n+2\ov 3n}} \pi g\ov l(n+1)}+{4 g^2 (n-1) \sqrt{{n+2\ov n}} \pi^2\ov 3l (n+1)},\cr
\G_{1,3}&=\G_{3,1}=0,\cr
\G_{2,2}&=\D_2-{2\sqrt 3 \pi (-4 +l \e  + l \e n^2 ) g\ov
 3l (n^2-1)}+{4 g^2 (n^2+3) \pi^2\ov 3l (n^2-1)},\cr
\G_{2,3}&=\G_{3,2}=-{(-2 + l\e) \sqrt{{n-2\ov 3n}} (n+1) \pi g\ov (n-1)}+{4g^2\sqrt{{n-2\ov n}}(n+1)\pi^2\ov 3l(n-1)},\cr
 \G_{3,3}&=\D_3+{(2 + l\e (n-2)) (n-3) \pi g\ov \sqrt 3 l(n-1)}+{8 g^2 \pi^2(n-2 )\ov 3l (n-1)}\ }
 $$
where
$$
\eqalign{
\D_1&=1 -{n+1\ov 2} \e + {l\ov 16} (n^2-1) \e^2,\quad
\D_2=1+{l\ov 16} (n^2-1) \e^2,\cr
\D_3&=1 +{n-1\ov 2} \e + {l\ov 16} (n^2-1) \e^2. }
$$
Evaluating this matrix at the fixed point \fx, we get
$$
\eqalign{
\G_{1,1}^{g^*}&=1 + {(20 - 4 n^2) \e\ov 8 (n+1)} + {l(39 - n - 7 n^2 + n^3) \e^2\ov
 16 (n+1)},\cr
\G_{1,2}^{g^*}&=\G_{2,1}^{g^*}={(n-1) \sqrt{{n+2\ov n}} \e(1+l\e)\ov n+1},\cr
\G_{1,3}^{g^*}&=\G_{3,1}^{g^*}=0,\cr
\G_{2,2}^{g^*}&=1 + {4 \e\ov n^2-1} + {l(65 - 2 n^2 + n^4) \e^2\ov 16 (n^2-1)},\cr
\G_{2,3}^{g^*}&=\G_{3,2}^{g^*}={\sqrt{{n-2\ov n}} (n+1) \e(1+l\e)\ov n-1},\cr
\G_{3,3}^{g^*}&=1 + {(n^2-5) \e\ov 2 (n-1)} + {l(-39 - n + 7 n^2 + n^3) \e^2\ov
 16 (n-1)} }
$$
whose eigenvalues are (up to order $\e^2$):
$$
\eqalign{
\D_1^{g^*}&=1 +  {1 + n\ov 2} \e + {l(7 +8 n + n^2)\ov 16} \e^2,\cr
\D_2^{g^*}&=1 + {l(n^2-1)\ov 16} \e^2,\cr
\D_3^{g^*}&=1 + {1-n\ov 2} \e +  {l(7 - 8 n + n^2)\ov 16} \e^2. }
$$

This result coincides with the dimensions $\tilde\D_{n+2,n}(l,p-l)$, $\D_{n,n}(l,p-l)+1$ and $\tilde\D_{n-2,n}(l,p-l)$ of the model $M(k-l,l)$ up to this order.
The corresponding normalized eigenvectors should be identified with the fields of $M(k-l,l)$:
$$
\eqalign{ \tphi_{n+2,n}(l,p-l)&={2 \ov n (n+1)}\phi_1^{g^*} + {2
\sqrt{{n+2\ov n}}\ov n+1}\phi_2^{g^*} + {\sqrt{n^2-4}\ov
n}\phi_3^{g^*},\cr
\phi_2(l,p-l)&=-{2 \sqrt{{n+2\ov n}}\ov n +
1}\phi_1^{g^*} -{n^2-5\ov n^2+1}\phi_2^{g^*} +{2\sqrt{{n-2\ov n}}\ov n-1}\phi_3^{g^*},\cr
\tphi_{n-2,n}(l,p-l)&={\sqrt{n^2-4}\ov n}\phi_1^{g^*}  - { 2
\sqrt{{n-2\ov n}}\ov n-1}\phi_2^{g^*} +{ 2\ov n(n-1)}\phi_3^{g^*}. }
$$
We used as before the notation $\tphi$ for the descendent field defined as in \defn\ and
$$
\phi_2(l,p-l)={1\ov 2\D_{n,n}^{p-l}(2\D_{n,n}^{p-l}+1)}\p\bar\p
\phi_{n,n}(l,p-l)
$$
is the normalized derivative of the corresponding primary field. We notice that these eigenvectors are finite
as $\e\rightarrow 0$ with exactly the same entries as in $l=1$ \pogt\ and  $l=2$ \as\ minimal models. This is one of the main results of this paper.

\newsec{Concluding remarks}

In conclusion, we considered here the RG flow of the general $\hat{su}(2)$ coset models $M(k,l)$ with $k\rightarrow\infty$ perturbed by the least relevant field up to the second order of the perturbation theory. As expected, we confirm that there is a nontrivial fixed point that coincides with the model $M(k-l,l)$ as was established time ago \myo. We defined certain descendent fields with dimension close to one and imposed conditions on their fusion rules so that they could mix along the RG flow. We explained how to compute the necessary structure constants and four-point functions. This is based on the construction of the the model of level $l$ as a projected tensor product of lower level models. We computed the anomalous dimensions of these fields which turn to coincide with the dimensions of certain fields of the model $M(k-l,l)$. The mixing matrix between these fields is then compared to the one recently considered in \mymy\ using the so called RG domain wall construction. We found an agreement between the two approaches. Moreover, these coefficients are finite, do not depend on $l$ and coincide with the corresponding coefficients for $l=1$ and $l=2$.

It will be interesting to examine the mixing of other fields, for example the analogs of the fields $\phi_{n,n\pm 1}$ in Virasoro and superconformal cases. It is clear that our method could be applied also for other coset models based on some algebra $\hat g_k$. The necessary ingredients for such calculations are the knowledge of the structure constants and the conformal blocks at just the first level.

\vskip1cm
\noindent
{\bf Acknowledgements}

\no This work was supported in part by the Bulgarian NSF Grant DFNI T02/6.

\appendix{A}{Structure constants}

We are interested in the mixing of the fields $\tphi_{n,n\pm 2}$ and the derivative $\p\bar\p\phi_{n,n}$ which were defined in \defn\ and \defnn. For that purpose we need to compute the structure constants involving these fields and the perturbing field $\tphi_{1,3}$. Following \myo\ we demand that the corresponding fusion rules are closed.
This requirement defines the coefficients in \defn\ and the corresponding structure constants. The latter govern the first order perturbative contribution and also enter the construction of the correlation functions we present in the next Appendixes.
So we impose the conditions:
\eqn\frn{
\eqalign{
\tilde\phi_{1,3}(l,p)\tilde\phi_{n,n+2}(l,p) &=\CC_{(13)(nn+2)}^{(nn)}\phi_{n,n}(l,p)+\CC_{(13)(nn+2)}^{(nn+2)}
\tilde\phi_{n,n+2}(l,p),\cr
\phi_{3,3}(l,p)\phi_{n,n}(l,p) &=\CC_{(33)(nn)}^{(nn+2)}\tilde\phi_{n,n+2}(l,p)+\CC_{(33)(nn)}^{(nn)}\phi_{n,n}(l,p)}}
and similarly for $\tilde\phi_{n,n-2}(l,p)$
As in \myo , using the explicit constructions of the fields we obtain functional equations for the coefficients and the structure constants \mymy. In order to solve these functional equations we use the fact that we know the value of the structure constants $\CC(1.p)$, i.e. the Virasoro ones. Also, by construction, the fields $\phi_{3,3}(l,p)$ and $\phi_{n,n}(l,p)$ are primary. Therefore their structure constants are just a product of lower level ones \mymy. Finally, one can use the knowledge of the solutions for $l=1,2,4$ \refs{\df ,\zp ,\pogtri}. With all this, we can make a guess and check it directly. Here is the list of the structure constants we will need:
\eqn\cnn{
\eqalign{
\CC_{(33)(nn)}^{(nn)}(l,p) &={\CG_n(p+l-1)\over\CG_n(p-1)},\cr
\CC_{(33)(nn)}^{(n+2n+2)}(l,p) &={\tilde\CG_n(p+l-1)\over\tilde\CG_n(p-1)},\cr
\CC_{(33)(nn)}^{(nn+2)}(l,p) &=\sqrt{{l\ov (p-n-1)(p+l-n-1)}}{\tilde\CG_n(p+l-1)\over\CG_n(p-1)},\cr
\CC_{(33)(nn+2)}^{(n+2n+2)}(l,p) &=-2\sqrt{{l\ov (p-n-1)(p+l-n-1)}}{\CG_{n+2}(p+l-1)\over\tilde\CG_n(p-1)}}}
\eqn\resn{
\eqalign{
\CC_{(13)(nn)}^{(nn)}(l,p) &=-(n-1)\sqrt{{l\ov (p+l-2)(p-2)}} \CG_n(p+l-1),\cr
\CC_{(13)(nn)}^{(nn+2)}(l,p) &=\sqrt{{(p+l-2)(p-n-1)\ov (p+l-n-1)(p-2)}} \tilde\CG_n(p+l-1),\cr
\CC_{(13)(nn+2)}^{(nn+2)}(l,p) &=\(-l(n+1)+{2(p+l-2)(p-n-1)\ov p+l-n-1}\){\CG_{n+2}(p+l-1)\ov \sqrt{l(p+l-2)(p-2)}},\cr
\CC_{(33)(nn+2)}^{(nn+2)}(l,p) &=(1-{2l\ov (p-n-1)(p+l-n-1)}){\CG_{-n+2}(p+l-1)\over\CG_{-n}(p-1)}
}}
where we introduced the functions
\eqn\gamn{
\eqalign{
\CG_n(p)&=\[\g^3({p\ov p+1})\g^2({2\ov p+1})\g^2({n-1\ov p+1})\g^2({p-n\ov p+1})\g({3\ov p+1})\]^{1\ov 4},\cr
\tilde\CG_n(p)&=\[\g({p\ov p+1})\g({n\ov p+1})\g({p-n-1\ov p+1})\g({3\ov p+1})\]^{1\ov 4}. }}
We want to stress that the "structure constants" thus obtained are actually square roots of the true structure constants $C$. The reason is that our construction makes use of "chiral" one-dimensional fields instead of the real two-dimensional ones \myt.
Therefore the true structure constants are squares of those in \cnn\ and \resn.

The coefficients in the construction \defn\ are given by:
$$
x=\sqrt{{(l-1)(p-n-1)\ov l(p-n)}} \qquad y=\sqrt{{p+l-n-1\ov l(p-n)}}.
$$

In exactly the same way one obtains the structure constants (and the coefficients $\tilde x$, $\tilde y$) involving the field $\tilde\phi_{n,n-2}(l,p)$. It turns out that they are obtained from the corresponding ones for $\tilde\phi_{n,n+2}(l,p)$ by simply changing $n\rightarrow -n$. This anticipates our observation in the main text that the two-point functions involving the field $\tilde\phi_{n,n-2}(l,p)$ are obtained from those of $\tilde\phi_{n,n+2}(l,p)$ by the same substitution.

Finally $\CC_{(13)(nn+2)}^{(nn-2)}(l,p)=0$ as can be seen by examining recursively the OPEs and fusion rules of the fields.

\appendix{B}{Correlation function $<\tphi(x)\tphi(0)\tphi(1)\tphi(\infty)>$}

In this Appendix we present the calculation of the correlation function
\eqn\fotri{
\eqalign{
&<\tphi(x)\tphi(0)\tphi(1)\tphi(\infty)>=\cr
&=<\prod_{i=1}^4\left(a(l,p)\phi_{1,1}(1,p)\tilde\phi_{1,3}(l-1,p+1)+b(l,p)\phi_{1,3}(1,p)\phi_{3,3}(l-1,p+1)\right)(x_i)>.}}

It defines the $\b$-function and the fixed point up to a second order of the perturbation theory. The correlation function is build out of conformal blocks.  Here we will use the construction presented in \myt. As explained in Chapter 3 the conformal blocks corresponding to \fotri\ are linear combinations of products of conformal blocks at levels $1$ and $l-1$ \cbprod. There are in general 16 terms in \fotri. Some of them are absent because of the fusion rules in each intermediate channel.
Here there are three channels: identity $\phi_{1,1}$, the field $\tilde\phi_{1,3}$ itself and $\tilde\phi_{1,5}$ which was defined in \five. We present the calculation of the corresponding conformal blocks separately.
Our strategy here is to compute the conformal blocks up to a sufficiently high order and to make a guess. For $l=1$ this was done in \pogt. For $l-1$ we proceed recursively and use the fact that we know the result for $l=2,3,5$. The calculations are simplified significantly by the fact that we need the result in the leading order in $\e\rightarrow 0$.

\vskip.5cm
\rb{\bf Channel $\phi_{1,1}$}

The possible internal channels in the product \cbprod\ are $r,t=1,n$ and $t,s=n,1$ with $n=1,3,5,...$ (odd integer) corresponding to descendants at higher level as in \exp. Let us examine the various terms that enter the sum \fotri\ and call for simplicity the corresponding conformal block at level $l$ $F_l$ omitting the indexes. The recursive calculation of the conformal blocks shows that they are finite at $\e\rightarrow 0$. Instead, some of them are multiplied (like in \cbprod) by a (square root of) the structure constant
$\CC_{(33)(33)}^{(31)}$ which is of order $\e^2$ (as can be seen from Appendix A). Therefore we can drop the corresponding terms. In addition, there is obviously a term where the identity multiplies the conformal block $F_{l-1}$ itself. A similar result comes from a product of $F_1$ with a conformal block of the field $\phi_{3,3}(l-1)$. The latter is equal to one at this order which is natural since the dimension $\D_{3,3}\sim 0$. Next, there are 4 terms which are simply two-point functions and therefore normalized to one (eventually multiplied by $\CC_{(13)(33)}^{(31)}$). An exception is given by terms where the fields $\phi_{1,3}(1)$ or $\tphi_{1,3}(l-1)$ are at positions $x$ and $1$. It can be shown that in this cases they give a contribution ${1\ov (1-x)^2}$ (since the corresponding dimensions are close to $1$).
We shall omit the detailed calculations which are straightforward but quite tedious. As a result of the above observations we get a recursive equation for the conformal block at level $l$ (omitting the obvious indexes):
\eqn\reco{
F_l=a^4F_{l-1}+b^4F_1+2a^2b^2+2a^2b^2 x^2 \CC_{(13)(33)}^{(31)}(l-1)\left(1+{1\ov (1-x)^2}\right)}
(note that we dropped the overall factor $x^{-2}$ for the time being).
The values of the coefficients in \reco\ in the leading order are:
$$
a\sim\sqrt{{l-1\ov l}},\qquad b\sim\sqrt{{1\ov l}},\qquad \CC_{(13)(33)}^{(31)}\sim {1\ov 3}.
$$
Also
$$
F_1={1\over (1-x)^2}(1-2x+3x^2-2x^3+1/3x^4)
$$
as computed in \pogt. Introducing the useful notation
$$
\tilde F_l=(1-x)^2F_l
$$
the recursion equation \reco\ becomes:
\eqn\rect{
l^2\tilde F_l=(l-1)^2\tilde F_{l-1}+\tilde F_1+2(l-1)f(x)}
where we defined
\eqn\fx{
f(x)=(1-x)^2+{x^2\ov 3}(1+(1-x)^2).}
The solution of this equation is given by:
$$
\tilde F_l={1\ov l}\tilde F_1+{l-1\ov l}f(x)
$$

Inserting $f(x)$ and returning to the initial notations (and restoring the overall $x^{-2}$) we get the final result for the conformal block:
\eqn\cbot{
\eqalign{
&<\tphi_{1,3}(x)\tphi_{1,3}(0)|_{11}\tphi_{1,3}(1)\tphi_{1,3}(\infty)>=\cr
&={1\ov x^2(1-x)^2}\left[1-2x+({5\ov 3}+{4\ov 3l})x^2-({2\ov 3}+{4\ov 3l})x^3+{1\ov 3}x^4\right].}}

This result is in perfect agreement with $l=1$ \pogt\ and $l=2$ \as.
\vskip.5cm
\rb{\bf Channel $\tphi_{1,5}$}

The field $\tphi_{1,5}$ was defined in \five\ and has a dimension close to 2. Therefore the possible internal channels in the product \cbprod\ are $r,t=1,n$ and $t,s=n,5$ with $n=1,3,5,...$.
Following the same logic as before we notice that the conformal blocks are finite but some of them are multiplied by (square roots of) structure constants that tend to zero in this order:
$$
\CC_{(33)(33)}^{(15)}\sim \e^4,\qquad \CC_{(33)(33)}^{(35)}\sim \e^2
$$
as can be seen from the explicit values in the previous Appendix and \confive. The analysis goes in the same way as before: there is a term proportional to $\CC_{(13)(13)}^{(15)}(l-1)F_{l-1}$ and similarly $\CC_{(13)(13)}^{(15)}(1)F_{1}$. The structure constants appear because of our construction \cbprod. Because of that we shall get as a result the conformal block multiplied by the corresponding structure constant
$\hat F_l=\CC_{(13)(13)}^{(15)}(l)F_{l}$. There are also two "free terms" multiplied by the constant
$$
\CC_{(13)(33)}^{(35)}(l-1)\sim \tilde G_3(p+l-1)\sim {\sqrt{5}\ov 3}
$$
in the leading order. In the same way as above we get a recursive equation for $\hat F_l$:
$$
l^2\hat F_l=(l-1)^2\hat F_{l-1}+x^2\hat F_1+2(l-1){\sqrt{5}\ov 3}\left(1+{1\ov (1-x)^2}\right).
$$
Denoting again $\tilde F_l=(1-x)^2\hat F_l$ we obtain:
$$
l^2\tilde F_l=(l-1)^2\tilde F_{l-1}+x^2\tilde F_1+2(l-1)f(x)
$$
where now
$$
f(x)={\sqrt{5}\ov 3}(1+(1-x)^2)={\sqrt{5}\ov 3}(2-2x+x^2).
$$
The solution of this equation is similar to the one we obtained above:
$$
\tilde F_l={1\ov l}x^2\tilde F_1+{l-1\ov l}f(x).
$$
Now we use the fact that we know the conformal block $\tilde F_1={\sqrt{5}\ov 3}$ from \pogt. Restoring the initial notations we obtain for the conformal block with internal channel $\tphi_{1,5}$ (note that here the overall power of $x$ is simply $x^0=1$):
\eqn\cbof{
\eqalign{
&<\tphi_{1,3}(x)\tphi_{1,3}(0)|_{15}\tphi_{1,3}(1)\tphi_{1,3}(\infty)>=\cr
&={1\ov (1-x)^2}\left[1-x+{l\ov 2(l-1)}x^2\right].}}
For the corresponding structure constant in the leading order we get:
$$
\CC_{(13)(13)}^{(15)}(l)\sim{2(l-1)\ov l}{\sqrt{5}\ov 3}.
$$
The same value one can obtain directly from \confive . All these results are in a perfect agrement with $l=1$ \pogt\ and $l=2$ \as.

\vskip.5cm
\rb{\bf Channel $\tphi_{1,3}$}

One can proceed in the same way as for the previous channels. It turns out however that in this case some of the conformal blocks that enter the sum \fotri are divergent as $p\rightarrow\infty$. These divergences  are exactly compensated by the zeros of the corresponding structure constants in \cbprod. Since the analysis similar to the above channels is more complicated here we adopt another strategy. Namely, we use the crossing symmetry of the correlation function \fotri.  We ask that it is invariant under the transformation \infi\ and use the explicit form of the remaining conformal blocks that we obtained above. This leads to linear equations for the coefficients in the $x$-expansion of the desired conformal block. The result is:
\eqn\cbthr{
\eqalign{
&<\tphi_{1,3}(x)\tphi_{1,3}(0)|_{13}\tphi_{1,3}(1)\tphi_{1,3}(\infty)>=\cr
&={1\ov x(1-x)^2}\left[1-{3\ov 2}x+{l+1\ov 2}x^2-{l\ov 4}x^3\right].}}

Combining altogether we finally obtain the 2D correlation function:
\eqn\fper{
\eqalign{
<\tphi(x)\tphi(0)\tphi(1)\tphi(\infty)>&=\left|{1\ov x^2(1-x)^2}\left[1-2x+({5\ov 3}+{4\ov 3l})x^2-({2\ov 3}+{4\ov 3l})x^3+{1\ov 3}x^4\right]\right|^2+\cr
&+{16\ov 3l^2}\left|{1\ov x(1-x)^2}\left[1-{3\ov 2}x+{l+1\ov 2}x^2-{l\ov 4}x^3\right]\right|^2+\cr
&+{5\ov 9}\left({2(l-1)\ov l}\right)^2\left|{1\ov (1-x)^2}\left[1-x+{l\ov 2(l-1)}x^2\right]\right|^2.}}

We used this function in Chapter 4 for the computation of the $\b$-function and the fixed point.

\appendix{C}{Other correlation functions}

In this Appendix we present the calculation of the other correlation functions we used in Chapter 6 to describe the mixing of the fields.

First, we notice that the computation of the function $<\tphi(x)\tphi(0)\tphi_{n,n+2}(1)\tphi_{n,n+2}(\infty)>$ goes in the same way as that of the function of the perturbing field itself, the latter being just a special case $n=1$. There are again the same three internal channels. It turns out that the corresponding conformal blocks are exactly the same, in agreement with $l=1$ and $l=2$ cases. The difference is only in the structure constants. Omitting the details we present the final result:
\eqn\crfn{
\eqalign{
&<\tphi(x)\tphi(0)\tphi_{n,n+2}(1)\tphi_{n,n+2}(\infty)>=\cr
&=\left|{1\ov x^2(1-x)^2}\left[1-2x+({5\ov 3}+{4\ov 3l})x^2-({2\ov 3}+{4\ov 3l})x^3+{1\ov 3}x^4\right]\right|^2+\cr
&+{8\ov 3l^2}{n+3\ov n+1}\left|{1\ov x(1-x)^2}\left[1-{3\ov 2}x+{l+1\ov 2}x^2-{l\ov 4}x^3\right]\right|^2+\cr
&+\left({2(l-1)\ov l}\right)^2{(n+3)(n+4)\ov 18 n(n+1)}\left|{1\ov (1-x)^2}\left(1-x+{l\ov 2(l-1)}x^2\right)\right|^2.}}

\vskip.5cm
\rb{\bf Function $<\tphi(x)\tphi(0)\tphi_{n,n+2}(1)\tphi_{n,n-2}(\infty)>$}

The only internal channel in this function correspond to the field $\tphi_{1,5}$. As above we drop among the terms in \cbprod\ those proportional to $\CC_{(nn)(nn)}^{(13)}$, $\CC_{(33)(33)}^{(35)}$, $\CC_{(nn)(nn)}^{(15)}$ which are of order $\e$ or higher. There are again terms $F_{l-1}$ and $F_1$ multiplied by the corresponding structure constants as well as "free terms" proportional to identity and $(1-x)^{-2}$. Denoting again $\hat F_l=C_l F_l$ we obtain explicitly the recursion equation:
$$
\hat F_l={(l-1)^2\ov l^2}\hat F_{l-1}+{1\ov l^2}x^2\hat F_1+2{(l-1)\ov l^2}\CC f(x)
$$
where
$$
f(x)=1+{1\ov (1-x)^2}
$$
and
\eqn\cnot{
\eqalign{
\CC={\sqrt {5}\ov 3}&\big(\sqrt{\sqrt{{n-2\ov 3n}}\CC_{(nn+2)(n-2n-2)}^{(35)}(l-1)}+\cr
&+\sqrt{\sqrt{{n+2\ov 3n}}\CC_{(nn-2)(n+2n+2)}^{(35)}(l-1)}\big).}}

The structure constants in \cnot are not among those listed in Appendix A. We computed them up to a sufficiently high level and found that in the leading order in $\e$ they are independent of $l$. The result is:
$$
\CC=\sqrt{{1\ov 3n}\sqrt{n^2-4}}
$$
which turns to coincide exactly with $\CC_1$. The solution of the equation then in terms of $\tilde F_l=(1-x)^2\hat F_l$ is:
$$
\tilde F=\CC\left( {2(l-1)\ov l}- {2(l-1)\ov l}x+x^2\right).
$$
Returning back to the original notations we can write finally the result for the correlation function:
\eqn\cfpm{
\eqalign{
&<\tphi(x)\tphi(0)\tphi_{n,n+2}(1)\tphi_{n,n-2}(\infty)>=\cr
&={1\ov 3n}\sqrt{n^2-4}\left( {2(l-1)\ov l}\right)^2
\left|{1\ov (1-x)^2}\left(1-x+{l\ov 2(l-1)}x^2\right)\right|^2.}}

\vskip.5cm
\rb{\bf Function $<\tphi(x)\tphi(0)\phi_{n,n}(1)\tphi_{n,n+2}(\infty)>$}

There is only one relevant intermediate internal channel in the leading order in this function corresponding to $\tphi_{1,3}$.
Inserting the definitions of the descendent fields we get seven terms in the sum representing the corresponding conformal block in \fotri. We note that there are again terms proportional to (square roots of) $\CC_{(nn)(nn)}^{(13)}$, $\CC_{(nn)(nn)}^{(15)}$, $\CC_{(33)(33)}^{(53)}$ which are of order $\e$ or higher. Only two terms remain in the conformal block:
$$
\eqalign{
&<\tphi(x)\tphi(0)|_{13}\tphi_{n,n}(1)\tphi_{n,n+2}(\infty)>=x^{-1}\big(a^2x\sqrt{\CC_{(13)(13)}^{(13)}(l-1)\CC_{(nn)(nn+2)}^{(13)}(l-1)}+\cr
&+b^2y\sqrt{\CC_{(13)(13)}^{(13)}(1)\CC_{(nn)(nn+2)}^{(13)}(1)\CC_{(33)(33)}^{(33)}(l-1)\CC_{(33)(nn)}^{(n+2n+2)}(l-1)}\big).}
$$
Inserting the values of the structure constants from Appendix A in the leading order we get finally for the correlation function:
\eqn\cfnpl{
<\tphi(x)\tphi(0)\tphi_{n,n}(1)\tphi_{n,n+2}(\infty)>={4\ov 3l}\sqrt{{n+2\ov n}}|x|^{-2}.}

\vskip.5cm
\rb{\bf Function $<\tphi(x)\tphi(0)\phi_{n,n}(1)\phi_{n,n}(\infty)>$}

It was mentioned in  Chapter 5 that this correlation function is exactly equal to those of $l=1$ and $l=2$. Here we want to explain in more details what is the reason for that. Note that, as mentioned in \pogt, we have to keep terms up to order $\e^2$ in this correlation function. Since the correlation function is quadratic in the conformal blocks we keep in the latter only terms up to order $\e$.

Since $\phi_{n,n}$ is just a primary field only four terms appear in this correlation function.
There are two relevant contributions in the intermediate channels corresponding to $\phi_{1,1}$ and $\tphi_{1,3}$.

Let us consider first the contribution from $\phi_{1,1}$.
There are two terms proportional to square roots of products of the constants $\CC_{(nn)(nn)}^{(13)}\CC_{(nn)(nn)}^{(31)}$  which are of order $\e^4$. As explained above we drop them.  Inserting the values of the structure constants in the leading order in the remaining two terms gives:
$$
F_l=a^2F_{l-1}+b^2F_1={l-1\ov l}F_{l-1}+{1\ov l}F_1.
$$
This equation is easily solved recursively:
$$
F_l=F_1
$$
(where $F_1$ is that of \pogt).

Similarly, in the channel corresponding to $\tphi_{1,3}$ there remain two terms, the other being of order $\e^2$
so we drop them. Then the equation reads:
$$
\eqalign{
\hat F_l&=a^2 F_{l-1}\sqrt{\CC_{(13)(13)}^{(13)}(l-1)\CC_{(nn)(nn)}^{(13)}(l-1)}+\cr
&+b^2 F_{1}\sqrt{\CC_{(13)(13)}^{(13)}(1)\CC_{(nn)(nn)}^{(13)}(1)\CC_{(33)(33)}^{(33)}(l-1)\CC_{(33)(nn)}^{(nn)}(l-1)}=\cr
&=\sqrt{{2(n^2-1)\ov 3p^2}}\left(a^2F_{l-1}+b^2F_1\right)}
$$
and we inserted the values of the structure constants. We see that the overall constant do not depend on $l$ so that this equation is very similar to the previous one and again the solution is:
$$
\hat F_l=\hat F_1=\sqrt{{2(n^2-1)\ov 3p^2}}F_1.
$$

As a result the correlation function is the same for all $l$ and reads (up to order $\e^2$):
\eqn\cfnn{
\eqalign{
&<\tphi(x)\tphi(0)\tphi_{n,n}(1)\tphi_{n,n}(\infty)>=\left|F_1(1,1)\right|^2+{2(n^2-1)\ov 3p^2}\left|F_1(1,3)\right|^2=\cr
&=|x|^{-4}+{(n^2-1)\e^2\ov 12}|x|^{-4}\left({x^2\ov 2(1-x)}+{\bar x^2\ov 2(1-\bar x)}+(\log (1-x)+\log (1-\bar x))^2\right).}}

\listrefs

\bye